\documentclass{aa}
\usepackage{graphicx}
\usepackage{txfonts}
\usepackage{natbib}
\usepackage{verbatim}
\usepackage{longtable}
\usepackage{lscape}
\usepackage{afterpage}
\usepackage{multirow}
\bibpunct{(}{)}{;}{a}{}{,}

\begin{document}

\title{Diagnostics for \emph{specific} PAHs in the far\textendash IR: searching
neutral naphthalene and anthracene in the Red Rectangle}

\author{G. Mulas\inst{1,2}
	\and
	G. Malloci\inst{2}
	\and
	C. Joblin\inst{2}
	\and
	D. Toublanc\inst{2}
}


\institute{INAF~\textendash~Osservatorio Astronomico di Cagliari~\textendash~Astrochemistry 
Group, Strada 54, Loc. Poggio dei Pini, I\textendash09012 Capoterra (CA), Italy \textemdash{}
\email{gmulas@ca.astro.it}
\and
Centre d'Etude Spatiale des Rayonnements, CNRS et Universit\'e Paul 
Sabatier\textendash Toulouse~3, Observatoire Midi-Pyr\'en\'ees, 9 Avenue du Colonel Roche,
31028 Toulouse Cedex 04, France \textemdash{}
\email{[giuliano.malloci; christine.joblin; dominique.toublanc]@cesr.fr}
}

\date{Received ?; accepted ?}

\abstract
{\emph{Context}. In the framework of the interstellar polycyclic aromatic hydrocarbons 
(PAHs) hypothesis, far\textendash IR 
skeletal bands are expected to be a fingerprint of single species in 
this class. \\
\emph{Aims}. We address the question of detectability of low energy PAH vibrational 
bands, with respect to spectral contrast and intensity ratio with ``classical''
Aromatic Infrared Bands (AIBs). \\
\emph{Methods}. We extend our extablished Monte\textendash Carlo model of the photophysics of 
specific PAHs in astronomical environments, to include rotational and
anharmonic band structure. The required molecular parameters were
calculated in the framework of the Density Functional Theory. \\
\emph{Results}. We calculate the detailed spectral profiles of three low\textendash energy vibrational
bands of neutral naphthalene, and four low\textendash energy vibrational bands of 
neutral anthracene.
They are used to establish detectability constraints based on 
intensity ratios with ``classical'' AIBs. A general procedure is suggested to
select promising diagnostics, and tested on available Infrared Space 
Observatory data for the Red Rectangle nebula. \\
\emph{Conclusions}. The search for single, specific PAHs in the far\textendash IR is a challenging, 
but promising task, especially in view of the forthcoming launch of the 
Herschel Space Observatory.
\keywords{Astrochemistry \textemdash{} Line: identification 
\textemdash{} Molecular processes \textemdash{} 
ISM: individual objects: Red Rectangle \textemdash{} 
ISM: lines and bands \textemdash ISM: molecules }}

\authorrunning{Mulas, Malloci, Joblin \& Toublanc}
\titlerunning{Searching neutral naphthalene and anthracene in the Red 
Rectangle}

\maketitle

\section{Introduction}\label{introduction}
The presence of polycyclic aromatic hydrocarbons (PAHs) in the 
interstellar medium (ISM) was proposed by \citet{leg84} and \citet{all85},
to account for the so\textendash called ``Aromatic Infrared Bands'' (AIBs), a set of 
emission bands observed near 3.3, 6.2, 7.7, 8.6, 11.3 and 12.7~$\mu$m, in many 
dusty environments excited by UV photons \citep{leg89,all89}. 
Such ``classical'' AIBs do not permit an unambiguous identification of 
any single PAH, since they just probe specific chemical bonds and not 
its overall structure. Indeed, despite the 
impressive amount of work devoted to this subject over the years, 
no \emph{definitive} spectral identification of any 
\emph{specific} individual member in this class exists to date.
On the other hand, every single such PAH ought to show a unique 
spectral fingerprint in the far\textendash IR spectral region, which contains 
the low\textendash frequency vibrational modes associated with collective oscillations 
of the whole skeletal structure of the molecule 
\citep{zha96,lan96,sal99b,job02,mul03,mul06,mul06b}. 

In \citet{mul06b} we presented and validated a procedure to model 
in a systematic way the photophysics of PAHs in 
photon dominated regions. In that work, we computed the complete 
far\textendash IR emission spectrum of a sample of 20 molecules and their cations
in three radiation fields,
covering some typical astronomical environments in which AIBs are 
observed. We concluded that the main problem for the detection and 
identification of such bands is likely to be spectral confusion and 
poor contrast against a strong background continuum: perpendicular
bands, which are expected to display sharp Q branches, are much
favoured with respect to shallower parallel bands.

Our approach can also model molecular rotation and thus 
obtain the expected band rotational envelopes of both far\textendash IR emission bands
\citep{job02} and visible absorption bands \citep{mal03c}. It can also 
estimate the contributions of fundamental and hot bands, with their 
distribution of anharmonic shifts, which affect the expected band profile.
This makes modelling much heavier: if $\mathrm{J}_\mathrm{max}$ is the 
maximum angular momentum which is significantly populated, 
$\mathrm{J}_\mathrm{max}\left(\mathrm{J}_\mathrm{max}+1\right)$ 
is the number of rotational levels which must be traced, during the 
simulation, for each vibrational mode. Since for the molecules considered 
$\mathrm{J}_\mathrm{max}$ is a few hundred \citep{rou97,mul98}, 
this makes the simulation 
$\mathrm{J}_\mathrm{max}\left(\mathrm{J}_\mathrm{max}+1\right)~\sim~10^5$ times
more demanding from a computational point of view. Furthermore, it 
requires the knowledge of the effective rotational constants as a 
function of vibrational state and of the anharmonic correction for hot 
bands. 

To elucidate the role of spectral shape on band detectability, we 
present here, as a test case, a detailed calculation for three low\textendash frequency
bands of the simplest PAH, namely neutral naphthalene 
(C$_{10}$H$_{8}$), 
and for four low\textendash frequency
bands of the next larger PAH in the group of oligoacenes, i.~e. 
neutral anthracene. 
(C$_{14}$H$_{10}$). 
In particular, for naphthalene we selected the lowest energy a\textendash type, b\textendash type
and c\textendash type transition, whose fundamental wavelengths are calculated 
to be respectively at 15.82, 27.74 and 58.56~$\mu$m; for anthracene, we 
selected the lowest energy  a\textendash type (43.68~$\mu$m) and c\textendash type (21.26, 26.35 
and 110.23~$\mu$m) bands.

In the following Sect.~\ref{modelling} we present our modelling approach. 
Results are then presented in Sect.~\ref{results} and discussed in 
Sect.~\ref{discussion}.

\section{Modelling approach} \label{modelling}

Our Monte\textendash Carlo modelling procedure is described in detail elsewhere 
\citep{mul98,job02,mal03c,mul03,mul06,mul06b}, as well as the
basic molecular parameters it requires. We here use exactly the same 
photo\textendash absorption cross\textendash sections and vibrational analyses 
of neutral napthtalene and neutral anthracene
as in \citet{mul06b}, where their applicability and accuracy
is discussed in detail. Since this is a proof\textendash of\textendash concept
work, we restricted ourselves to one single well\textendash known environment, 
i.~e. the Red Rectangle (RR) halo, using the same radiation field (RF) 
previously adopted for it in \citet{mul06}, derived from the observational 
work of \citet{vij05}. 

We here additionally studied in detail the distribution of anharmonic
shifts arising from the superposition of a large number of hot bands 
on top of the fundamental band, in the stochastic process of PAH
relaxation via IR emission. Moreover, we also calculated detailed
rotational profiles obtained from statistical equilibrium.

To do this we needed the previously unavailable anharmonic correction 
to vibrational terms and rotational constants, and the Coriolis 
vibration\textendash rotation coupling, which we computed using the 
\textsc{Gaussian03} quantum chemistry package \citep{g03}.
This implements the Van Vleck perturbative approach to the above problem 
\citep[e.~g.][]{cla88}, which we applied in the framework of the Density
Functional Theory, with 
the exchange\textendash correlation functional B3LYP \citep{bec93,ste94} and the 
\mbox{4-31G} gaussian basis set \citep{fri84}. While basis set convergence 
is not achieved yet at this level of theory, it is good enough to yield 
sensible results. Indeed, benchmark
calculations showed that smaller basis sets are required to obtain
good perturbative corrections than are needed for equilibrium 
geometry properties and harmonic vibrational analyses \citep{bar04,bar05}.

The anharmonic vibrational analysis yields the second\textendash order anharmonic
corrections to the vibrational energy levels, which are represented,
in wavenumbers, by the formula 
\citep[see e.~g.][ and references therein]{bar05}:
\begin{displaymath} 
\mathrm{E}_\mathrm{vib}(n_1,\ldots, n_N) = \chi_0+\!\!
\sum_i\left(n_i\!+\!\frac{1}{2}\right)\,\overline{\omega_i} 
+\!\!\sum_{i,j\geq i} \chi_{ij}\,\left(n_i\!+\!\frac{1}{2}\right)
\left(n_j\!+\!\frac{1}{2}\right),
\end{displaymath}
where $\chi_0$ and $\chi_{ij}$ are the vibrational anharmonic constants as defined
in \citet{bar95} with the corrections reported in \citet{bar04}.
The values of $\chi_{ij}$ which we obtained are listed in 
Tables~\ref{chinaphthalene} and \ref{chianthracene} 
in the Appendix for the specific bands we considered here.

Accidental near degeneracies, when they occur, lead to additional 
corrections (i.~e. Fermi and Darling\textendash Dennison resonances), which are not
systematic and must thus be evaluated, if relevant, on a case by case basis. 
For the three specific benchmark bands which we modelled here
for neutral naphthalene, the levels from which emission is estimated to 
occur significantly (fundamental and first few hot bands) are unperturbed by 
such resonances. For bands for which they do occur, in the case of neutral
naphthalene and anthracene they are smaller than anharmonic corrections 
anyway, so that the qualitative conclusions we may draw from our test cases 
remain valid.

To leading order, selection rules are the same as in the harmonic 
approximation, i.~e. they connect levels in which a single IR\textendash active
mode changes by one quantum. The energy difference 
\begin{displaymath}
\Delta\mathrm{E}_k\left(n_1,\ldots,n_N\right) = 
\mathrm{E}_\mathrm{vib}\left(\ldots,n_k+1,\ldots\right)-
\mathrm{E}_\mathrm{vib}\left(\ldots,n_k,\ldots\right)
\end{displaymath}
for a transition in which only the $k^\mathrm{th}$ quantum number changes as
$n_k+1 \to n_k$ is hence given by
\begin{displaymath}
\Delta\mathrm{E}_k\left(n_1,\ldots,n_N\right) =
\overline{\omega_k} + 2\chi_{kk}\left(n_k+1\right) + 
\sum_{i\neq k}\chi_{ik} \left(n_i+\frac{1}{2}\right), 
\end{displaymath}
which may be rewritten as
\begin{displaymath}
\Delta\mathrm{E}_k\left(n_1,\ldots,n_N\right) = \Delta\mathrm{E}_k\left(0,\ldots,0\right) + 
2\chi_{kk}n_k+\sum_{i\neq k}\chi_{ik}n_i.
\end{displaymath}
This equation shows that anharmonicity produces a shift in the hot bands, 
with respect to the fundamental, which is proportional to the vibrational 
quantum numbers in the lower energy state involved in the transition. 
Since IR emission by PAHs results from a vibrational cascade, a given band 
will result in the superposition of the fundamental plus a possibly large 
number of hot bands, their contributions depending on the detailed 
statistics of the process. This in turn results in a distribution of 
anharmonic shifts, which convolves the pure (non\textendash thermal) rotational 
envelope of the band. Both effects are sampled by our Monte\textendash Carlo 
procedure. 

\begin{figure}

\includegraphics[width=\hsize]{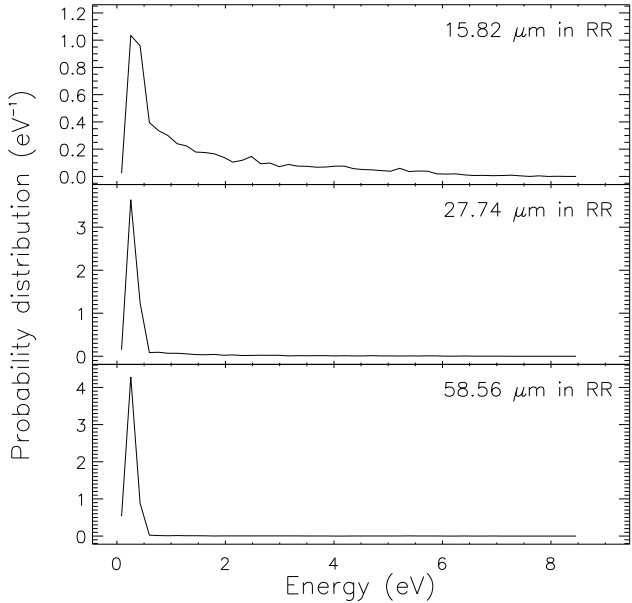}
\vspace{-0.4cm}
\caption{Probability distribution of the excitation energies at which photons 
are emitted in the three bands considered of neutral naphthalene in the RR 
halo.}
\label{naphthists}
\end{figure}

\begin{figure}

\includegraphics[width=\hsize]{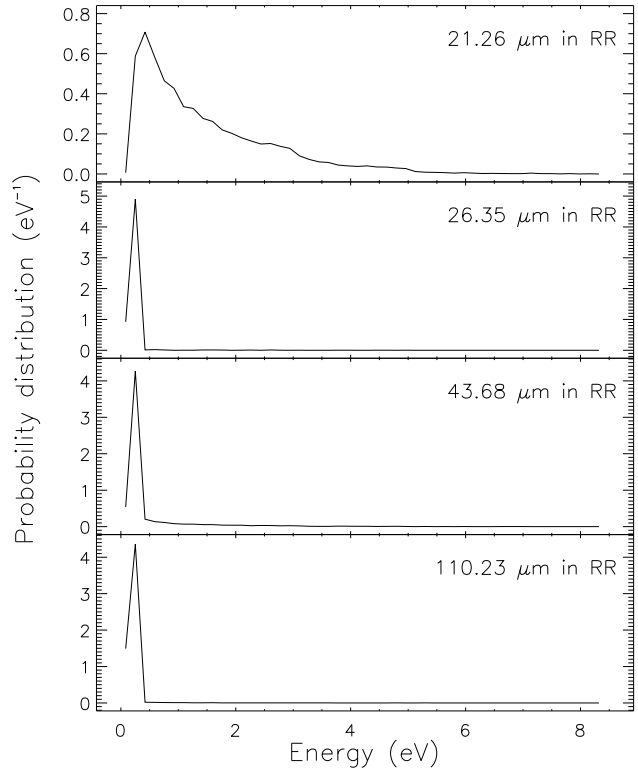}
\caption{Same as Fig.~\ref{naphthists} for the four bands considered of
neutral anthracene.}
\label{anthrhists}
\end{figure}

On top of this, each single transition, due to higher order perturbative 
effects, turns into the superposition of transitions including 
a distribution of nearby states. This produces a lorentzian envelope,
i.~e. lifetime broadening, with a width which roughly scales 
with the density of vibrational states at the energy at which the 
transition occurs. Since bands in the low\textendash energy modes are 
emitted near the end of the vibrational cascades (see Figs.~\ref{naphthists}
and \ref{anthrhists}), below or slightly above the decoupling energy 
\citep[we assumed E$_\mathrm{dec}~\simeq~0.32$~eV for neutral naphthalene and 
0.21~eV for neutral anthracene,][]{mul06b}, we neglect lifetime broadening 
in our present simulation. Specifically, two of the three bands we
modelled for naphthalene and three of the four for anthracene
are emitted virtually \emph{only} below E$_\mathrm{dec}$; only the highest 
energy band we considered for each molecule is emitted in a non negligible 
fraction at excitation energies above  E$_\mathrm{dec}$.
Moreover, comparison with experimental spectra of gas\textendash phase naphthalene 
at room temperature show a rather good correspondence of the band near 
$\sim$12.8~$\mu$m with a theoretical spectrum we calculated under these same 
assumptions \citep{pir06}.

As to the change of the effective rotational constants as a function of
vibrational state, to leading order they have a linear dependence on
vibrational quantum numbers, i.~e.
\begin{displaymath}
\mathrm{A}_\mathrm{eff}\left(n_1,\ldots,n_N\right) = 
\mathrm{A}_\mathrm{eff}\left(0,\ldots,0\right)-\sum_i\mathrm{a}_in_i
\end{displaymath}
and equivalent relations for $\mathrm{B}_\mathrm{eff}$ and 
$\mathrm{C}_\mathrm{eff}$. The $\mathrm{a}_i$,  $\mathrm{b}_i$ and 
$\mathrm{c}_i$ constants (the so\textendash called $\alpha$ matrix), which include three 
leading contributions from harmonic, anharmonic and Coriolis terms in the 
molecular Hamiltonian, are almost all very small for naphthalene and 
anthracene, producing changes in the range of a few parts in a thousand 
in the effective rotational constants. We report them in 
Tables~\ref{vibrot_alpha_naphthalene} and \ref{vibrot_alpha_anthracene} 
in the Appendix.

Only in two cases, for naphthalene, the accidental near resonance of the two 
vibrational modes respectively at 15.82 and 15.90~$\mu$m and at 19.49 and 
19.54~$\mu$m produces large Coriolis coupling terms between them, 
resulting in a $\sim4\%$ change in the affected rotational constant as a 
function of the number of quanta in that vibrational state. In the case of 
anthracene, the largest change in the rotational constants is less than
0.9\%, again due to an accidental near resonance between the modes at 25.45 
and 26.62~$\mu$m.

\section{Results} 
\label{results}
\subsection{Calculated band profiles}\label{profiles}

The three bands of neutral naphthalene for which we modelled the detailed 
rovibrational and anharmonic structure are one of each type, i.~e. an 
a\textendash type, a b\textendash type and a c\textendash type band respectively. They are two in\textendash plane 
and one out\textendash of\textendash plane (the latter being the so\textendash called ``butterfly'' or 
``floppying'') bending modes, and each of them is the lowest frequency 
transition of its type. For neutral anthracene we modelled the four 
lowest frequency bands, which are, in order of decreasing energy, two
c\textendash type, one a\textendash type and another c\textendash type.

\begin{figure}
 
\includegraphics[width=\hsize]{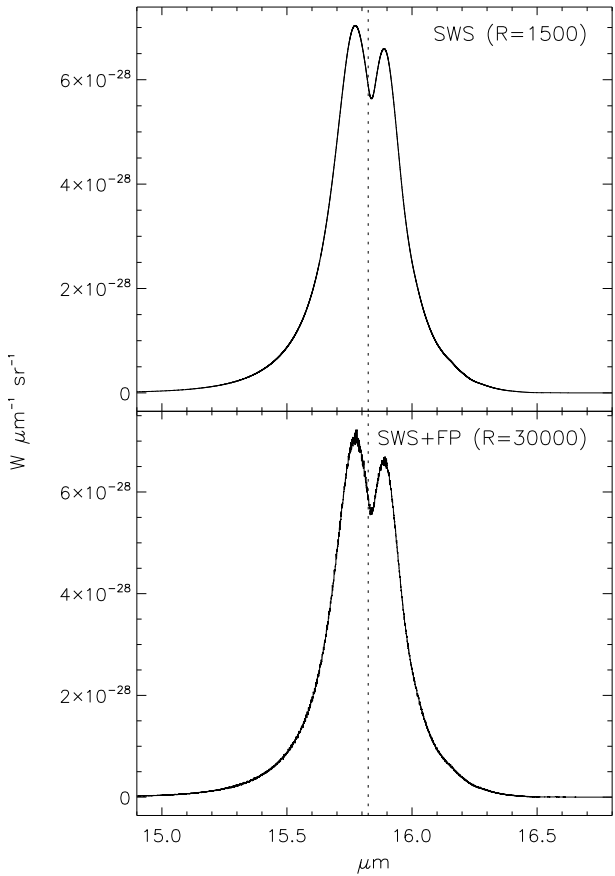}
\caption{Detailed rotational structure of the band at 
15.82~$\mu$m of neutral naphthalene in the RR halo. Strong Cariolis 
coupling with the IR\textendash inactive band at 15.90~$\mu$m causes one of the rotational
constants to differ by $\sim4\%$ between the upper and lower vibrational 
states involved in the transition, producing a blue shaded rotational
profile. The superposition of hot bands, which has a red\textendash shaded envelope, 
acts in the opposite direction, so that the overall profile including 
both effects is almost symmetric.
The two panels show the band as would be seen with ISO\textendash SWS with and without
the Fabry\textendash Perot filter. The vertical dotted line marks the position of the 
origin of the fundamental band.}
\label{naphtha_n_39}
\end{figure}

\begin{figure}
 
\includegraphics[width=\hsize]{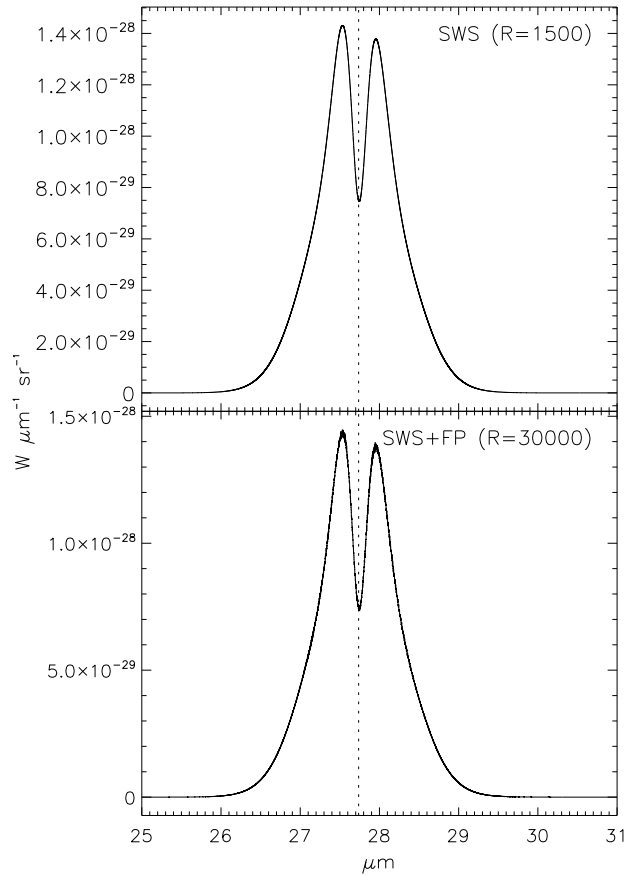}
\caption{Same as Fig.~\ref{naphtha_n_39} for the naphhalene band at 
27.74~$\mu$m. The rotational 
constants are almost unchanged in the vibrational transition, resulting in a 
very symmetric rotational envelope, the asymmetry being mainly due to the 
superposition of slightly displaced sovratones of the band.
}
\label{naphtha_n_46}
\end{figure}

\begin{figure}
 
\includegraphics[width=\hsize]{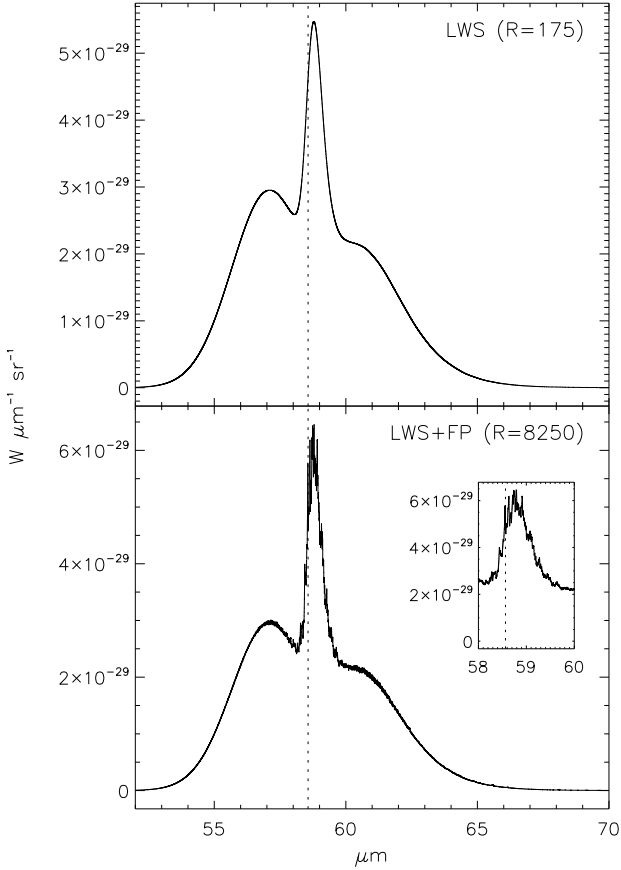}
\caption{Same as Fig.~\ref{naphtha_n_39} for the naphthalene band at 
58.56~$\mu$m. The rotational 
constants are almost unchanged in the vibrational transition, resulting in 
a very symmetric rotational envelope, with a Q branch standing out very 
clearly. The two panels show the band as would be seen with ISO\textendash LWS with and 
without the Fabry\textendash Perot filter. In the lower panel, the higher resolving 
power allows the resolution, in the Q branch, of the separate contributions 
of different hot bands, with slightly different anharmonic shifts (zoomed 
inset). 
}
\label{naphtha_n_48}
\end{figure}

\begin{figure}
 
\includegraphics[width=\hsize]{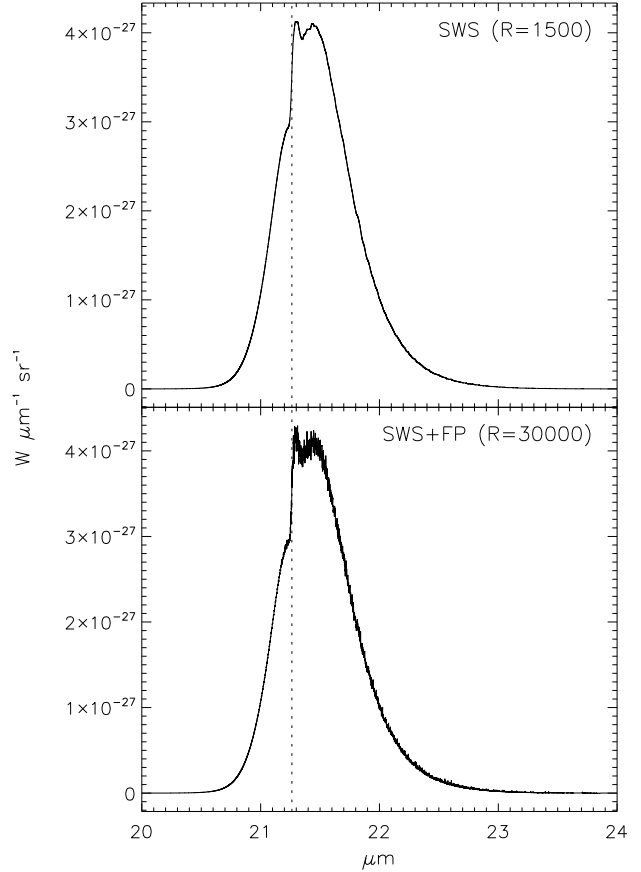}
\caption{Detailed rotational structure of the band at 
21.26~$\mu$m of neutral anthracene in the RR halo. The rotational 
constants are almost unchanged in the vibrational transition, resulting in 
a very symmetric rotational envelope, with a visible Q branch whose width is
due to the superposition of hot bands with different anharmonic shifts. 
The two panels show the band as would be seen with ISO\textendash SWS with and 
without the Fabry\textendash Perot filter. The vertical dotted lines marks the position 
of the origin of the fundamental band.
}
\label{anthra_n_58}
\end{figure}

\begin{figure}
 
\includegraphics[width=\hsize]{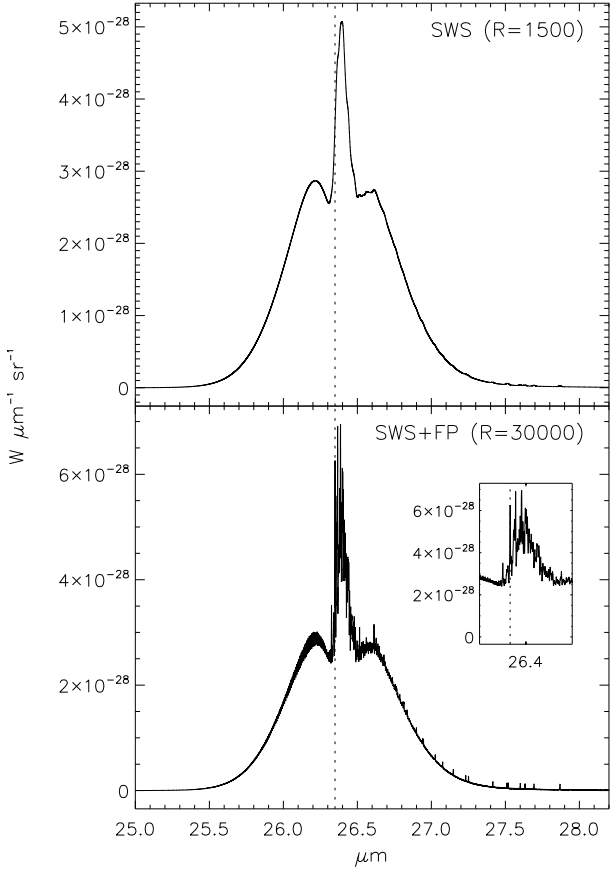}
\caption{Same as Fig.~\ref{anthra_n_58} for the anthracene band at 
26.35~$\mu$m. The rotational 
constants are almost unchanged in the vibrational transition, resulting in 
a very symmetric rotational envelope, with a sharp Q branch standing clearly
out in the middle. 
In the lower panel, the 
higher resolving power allows the resolution, in the Q branch, of the 
separate contributions of different hot bands, with slightly different 
anharmonic shifts (zoomed inset). 
}
\label{anthra_n_61}
\end{figure}

\begin{figure}
 
\includegraphics[width=\hsize]{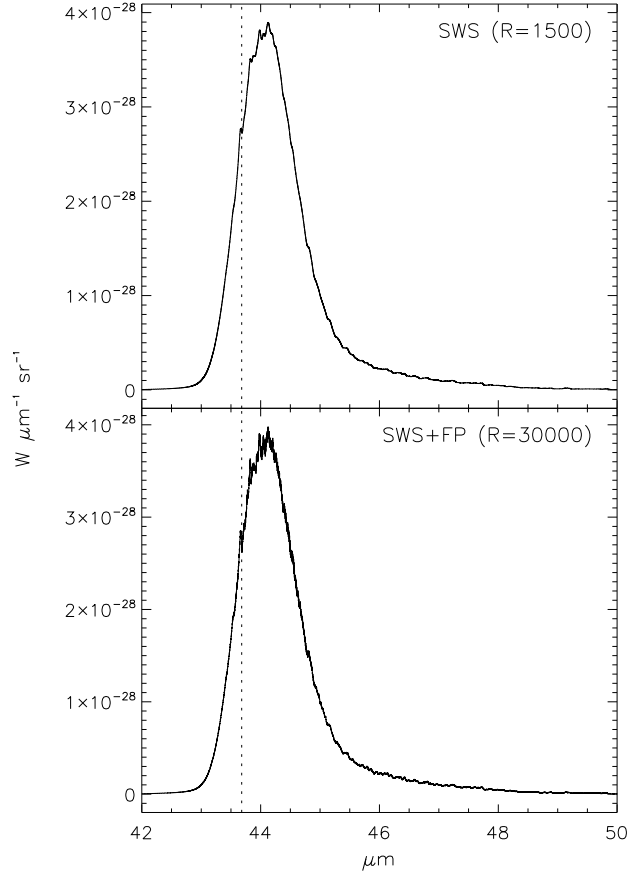}
\caption{Same as Fig.~\ref{anthra_n_58} for the anthracene band at 
43.68~$\mu$m. The relatively large 
change in the rotational constants in the vibrational transition results in 
a visibly red\textendash shaded band. 
}
\label{anthra_n_64}
\end{figure}

\begin{figure}
 
\includegraphics[width=\hsize]{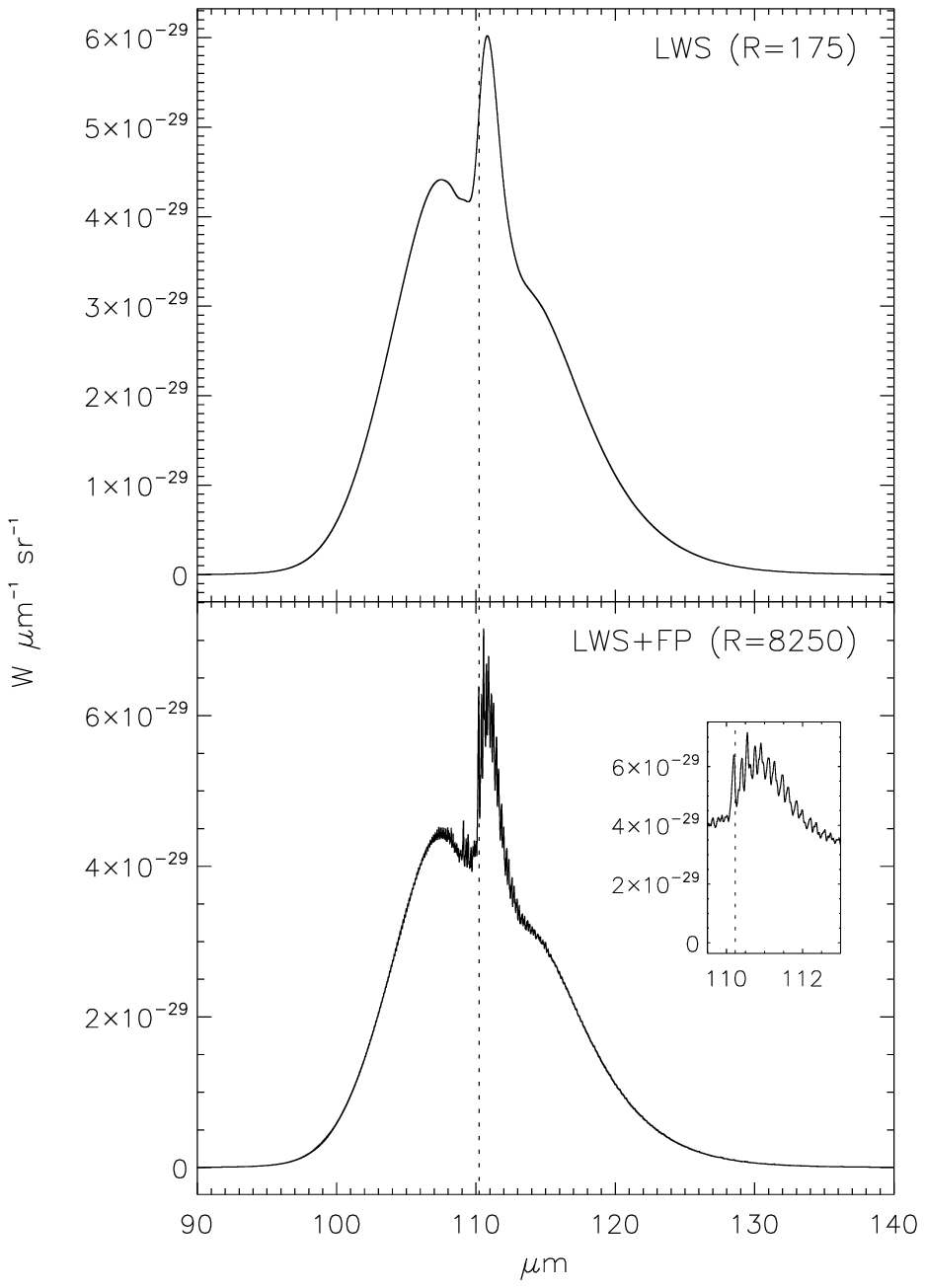}
\caption{Same as Fig.~\ref{anthra_n_58} for the anthracene band at 
110.23~$\mu$m. The change in rotational constants, combined with
the superposition of hot bands, produces an asymmetric profile, in the 
middle of which a sharp Q branch clearly stands out. In the lower panel, the 
higher resolving power allows the resolution, in the Q branch, of the 
separate contributions of different hot bands, with slightly different 
anharmonic shifts (zoomed inset).}
\label{anthra_n_66}
\end{figure}

Since the lowest energy butterfly mode of both molecules falls in the 
spectral range of the Long Wavelength Spectrograph (LWS) on board ISO,
we convolved their synthetic spectra with the average resolving power 
of LWS with (R$\simeq$8250) and without (R$\simeq$175) the Fabry\textendash Perot filter, 
assuming a velocity dispersion $\lesssim5$~km~s$^{-1}$. Similarly, since all
other bands fall in the spectral range covered by the Short Wavelength 
Spectrograph (SWS) of ISO, we convolved their synthetic spectra with 
the resolving power of SWS with (R$\simeq$30000) and without (R$\simeq$1500) 
the Fabry\textendash Perot filter, with the same velocity dispersion. 
For details on SWS and LWS, see their manuals on the official ISO web 
page\footnote{\texttt{http://www.iso.vilspa.esa.es}}.
The resulting spectra are shown in figures 
\ref{naphtha_n_39} to \ref{anthra_n_66}.


There is clearly a qualitative difference between perpendicular and parallel
bands, i.~e. the presence of a central Q branch containing about 20\% of the 
total flux in the band. Such difference is very apparent in the 
bands at 58.56~$\mu$m of naphthalene and at 26.35~$\mu$m of anthracene: in 
these cases, the rotational constants are very nearly the same in the two 
vibrational modes involved in the transition, which makes the resulting Q 
branches extremely sharp, with a full width at half maximum (FWHM) of the 
order of $\sim$0.15~cm$^{-1}$. Moreover, in the absence of 
shading\footnote{A difference in the rotational constants between the upper
and lower states involved in a transition leads to the inversion of some
branches, with the consequent formation of band heads, and results in an 
overall asymmetry in the rotational profile. A band is then
commonly said to be \emph{red\textendash shaded} when it has a more extended tail on 
the red side, \emph{blue\textendash shaded} in the opposite case \citep{her91a,her91b}},
neither the P nor the R branch is inverted, and the Q branch sits in the gap 
between them, with little contamination. The superposition of the fundamental 
and hot bands produces a well defined pattern of anharmonic shifts, clearly 
visible in the zoomed inset in the lower panel of Figs.~\ref{naphtha_n_48},
\ref{anthra_n_61} and \ref{anthra_n_66}. With the low resolution mode 
of LWS, the fundamental and hot bands merge in an unique, unresolved, 
red\textendash shaded Q branch, with a FWHM corresponding to the resolving 
power of the instrument. The P and R branches have a FWHM of the order of 
$\sim$8~cm$^{-1}$.

In the band at 21.26 of anthracene, instead, the larger
difference in rotational constants between the states involved in the
transition and the pattern of anharmonic shifts of the hot bands effectively
masks the central Q branch, which is just barely visible in the high 
resolution panel of Fig.~\ref{anthra_n_58}.

The parallel bands of naphthalene are resolved in only two well\textendash defined 
features, which in the b\textendash type one at 27.74~$\mu$m include a split Q branch
\citep{her91b}. The width of these partially resolved features is 
approximately $\sim$6~cm$^{-1}$. As to the parallel band of anthracene at 
43.68~$\mu$m, relatively strong red shading and anharmonicity combine
to produce a single, unresolved, asymmetric feature with an approximate
FWHM of $\sim$6~cm$^{-1}$.

These differences in band shapes, which depend on the interplay of 
many molecular parameters, have direct consequences on their 
detectability.

\subsection{Comparison with observations}\label{observations}

We here examine in detail the case of the RR nebula, 
which is one of the reference targets
for the observation of AIBs and thus a prime candidate for the 
identification of specific PAHs in space. ISO data for this object are
available in the online archive. 
Under optically thin conditions, 
the observed spectrum of a given band produced by a given PAH, for instance
neutral naphthalene, is
\begin{equation} \label{eq1}
\frac{dF}{d\lambda} = \frac{dF_\mathrm{cont}}{d\lambda} + 
\int dr\,d\Omega\,n_\mathrm{naph.}\frac{d\mathcal{P}}{d\lambda},
\end{equation}
where $\displaystyle \frac{dF_\mathrm{cont}}{d\lambda}$ is the underlying continuum 
spectrum, $\displaystyle \frac{d\mathcal{P}}{d\lambda}$ is the power isotropically 
emitted in the band by one molecule for the assumed RF, 
$n_\mathrm{naph.}$ is its number density and the double 
integral is along the line of sight and over the solid angle observed.
As explained in detail in \citet{mul06b}, the IR spectrum emitted by
a species in a given regime scales linearly with the intensity of the
RF. Since extinction in the RR halo is negligible \citep{vij05},
the RF will change only due to dilution with the position in the nebula
and such scaling relations are valid in this case.
Hence, we can factor $\displaystyle \frac{d\mathcal{P}}{d\lambda}$ 
as
\begin{displaymath}
\frac{d\mathcal{P}}{d\lambda} = 
\frac{d\mathcal{P}_\mathrm{ref}}{d\lambda}\,\Lambda\left(r,\Omega\right),
\end{displaymath}
where $\Lambda\left(r,\Omega\right)$ is an adimensional scaling factor, independent of 
$\lambda$, which takes into account the variation of RF intensity with position in 
the source and
$\displaystyle \frac{d\mathcal{P}_\mathrm{ref}}{d\lambda}$ is independent of $r$ and 
$\Omega$.
We may thus rewrite Eq.~(\ref{eq1}) as
\begin{eqnarray} \label{eq2} 
\frac{dF}{d\lambda} & = & \frac{dF_\mathrm{cont}}{d\lambda} + 
\frac{d\mathcal{P}_\mathrm{ref}}{d\lambda} 
\int dr\,d\Omega\,n_\mathrm{naph.}\,\Lambda\left(r,\Omega\right)\nonumber \\
& = & \frac{dF_\mathrm{cont}}{d\lambda}+
\frac{d\mathcal{P}_\mathrm{ref}}{d\lambda}\,\Upsilon_\mathrm{naph.},
\end{eqnarray}
where $\Upsilon_\mathrm{naph.}$, defined above, has the dimensions of a column density
times a solid angle and is independent of $\lambda$. Classical AIBs are commonly 
considered as typical tracers of PAHs. Using Eq.~(\ref{eq2}) and 
integrating over one AIB, we obtain
\begin{eqnarray}
\int_\mathrm{AIB}d\lambda\left(\frac{dF}{d\lambda}-\frac{dF_\mathrm{cont}}{d\lambda}\right) & = &
\Upsilon_\mathrm{naph.} \int_\mathrm{AIB}d\lambda \frac{d\mathcal{P}_\mathrm{ref}}{d\lambda} \nonumber \\
& = & \Upsilon_\mathrm{naph.}\,\,\mathcal{P}_\mathrm{AIB} \nonumber =  
\eta_\mathrm{AIB}\,F_\mathrm{AIB},\nonumber
\end{eqnarray}
where $\eta_\mathrm{AIB}$ is the fraction of the flux in the AIB produced by 
the given molecule and $\displaystyle \mathcal{P}_\mathrm{AIB}$ is the result 
of our model for the reference RF and the chosen band. Solving the above 
equation for $\Upsilon_\mathrm{naph.}$, we get
\begin{equation} \label{eq3}
\Upsilon_\mathrm{naph.} = \eta_\mathrm{AIB}\,\,\frac{F_\mathrm{AIB}}{\mathcal{P}_\mathrm{AIB}}.
\end{equation}
Let's now focus on another IR band of the same specific PAH. To be detectable,
it must stand above the continuum by more than the noise level, i.~e., for
a $2\textendash\sigma$ detection:
\begin{displaymath}
\frac{\displaystyle \frac{dF_\mathrm{peak}}{d\lambda}-\frac{dF_\mathrm{cont}}{d\lambda}}
{\displaystyle \frac{dF_\mathrm{cont}}{d\lambda}} \gtrsim \frac{2}{S},
\end{displaymath}
$S$ being the signal to noise ratio, from which follows, using Eq.~(\ref{eq2})
and then Eq.~(\ref{eq3}),
\begin{displaymath}
\frac{\displaystyle\left(\frac{d\mathcal{P}_\mathrm{ref}}{d\lambda}\right)_\mathrm{peak}
\Upsilon_\mathrm{naph.}} {\displaystyle\frac{dF_\mathrm{cont}}{d\lambda}} =
\frac{\displaystyle\left(\frac{d\mathcal{P}_\mathrm{ref}}{d\lambda}\right)_\mathrm{peak}
\eta_\mathrm{AIB}\,\,\frac{F_\mathrm{AIB}}{\mathcal{P}_\mathrm{AIB}}}
{\displaystyle\frac{dF_\mathrm{cont}}{d\lambda}} \gtrsim \frac{2}{S}.
\end{displaymath}
In the equation above, we implicitly assumed both this latter band and
the reference AIB are integrated over the same aperture
on the sky. For ISO observations of the RR, this is a good assumption since
RR is completely contained both in the SWS and in the LWS entrance 
apertures \citep{bre03,men02}.
Solving for $\eta_\mathrm{AIB}$, we finally obtain that the band will be detected 
only if
\begin{equation} \label{etalimit}
\eta_\mathrm{AIB} \gtrsim \frac{2}{S}\,\,\frac{\displaystyle\frac{dF_\mathrm{cont}}{d\lambda}}
{\displaystyle\left(\frac{d\mathcal{P}_\mathrm{ref}}{d\lambda}\right)_\mathrm{peak}}
\,\,\frac{\mathcal{P}_\mathrm{AIB}}{F_\mathrm{AIB}}.
\end{equation}
This derivation is completely general, despite the fact that we wrote
it for neutral naphthalene and applies to any specific band of any specific
PAH. We can now put the numbers for the three bands we studied in 
detail for neutral naphthalene. Upon examination of the previously
calculated spectra, the strongest emission bands of neutral naphthalene 
in the RR are the in\textendash plane C\textendash H stretch at $\sim3.3~\mu$m and the 
out\textendash of\textendash plane C\textendash H bend at $\sim12.7~\mu$m; the fraction of the total IR emission
of neutral naphthalene in such bands is calculated to be respectively 58.9\% 
and 26.5\%.
From observations of the RR available from the online ISO database, their 
integrated band intensities amount to 
$F_\mathrm{3.3}$~=~7.35~10$^{-17}$~W~cm$^{-2}$ and 
$F_\mathrm{12.7}$~=~3.51~10$^{-17}$~W~cm$^{-2}$. 
From our model we have, for neutral naphthalene,
$\mathcal{P}_\mathrm{3.3}$~=~2.42~10$^{-26}$~W~sr$^{-1}$ and
$\mathcal{P}_\mathrm{12.7}$~=~9.08~10$^{-27}$~W~sr$^{-1}$, while the values
of S and $\displaystyle\frac{dF_\mathrm{cont}}{d\lambda}$ estimated from ISO 
archive data and those of 
$\displaystyle\left(\frac{d\mathcal{P}_\mathrm{ref}}{d\lambda}\right)_\mathrm{peak}$
calculated by our model (see Figs.~\ref{naphtha_n_39} to \ref{naphtha_n_48})
are listed in Table~\ref{pupamela}, along with the resulting detection 
limits at $\sim$2$\sigma$ level. Since $\eta_\mathrm{AIB}$ was defined
as the fraction of the flux observed in a given AIB which is produced by 
this specific molecule, a detection limit $\eta_\mathrm{AIB}~>~1$ means that 
the corresponding band (1$^\mathrm{st}$ column in the table) is 
undetectable with the achieved S/N ratio.
\begin{table}
\caption{Detection limits estimated for three test bands of neutral 
naphthalene, along with the parameters used to derive them. Two entries
are listed for the perpendicular band at 58.56~$\mu$m, corresponding to
the resolving power of LWS with and without the Fabry\textendash Perot filter.}
\label{pupamela}
\begin{center}
\begin{tabular}{ccccccc}
\hline \hline \noalign{\smallskip} 
Band & S & $\displaystyle\frac{dF_\mathrm{cont}}{d\lambda}$ & 
R &
$\displaystyle\left(\frac{d\mathcal{P}_\mathrm{ref}}{d\lambda}\right)_\mathrm{peak}$ &
$\eta_\mathrm{3.3}$ & $\eta_\mathrm{12.7}$\\
\noalign{\smallskip} 
($\mu$m) & & $\displaystyle\left(\frac{\mathrm{W}}{\mathrm{cm^2~\mu m}}\right)$ & 
& $\displaystyle\left(\frac{\mathrm{W}}{\mathrm{sr~\mu m}}\right)$ & \\
\noalign{\smallskip} \hline \noalign{\smallskip}
15.82 & 25 & 4.6~10$^{-16}$ & \textemdash{} & 7.3~10$^{-28}$ & $>1$ & $>1$ \\
27.74 & 45 & 1.5~10$^{-16}$ & \textemdash{} & 1.5~10$^{-28}$ & $>1$ & $>1$ \\
\multirow{2}*{58.56} & \multirow{2}*{35} & \multirow{2}*{1.5~10$^{-18}$} &
l & 5.4~10$^{-29}$ & 0.52 & 0.42 \\
 & & & h & 6.5~10$^{-29}$ & 0.44 & 0.34\\
\noalign{\smallskip} \hline 
\end{tabular}
\end{center}
\end{table}
The perpendicular band, with its prominent central Q branch, is a much more 
sensitive (about two orders of magnitude) probe for neutral naphthalene, 
yielding the lowest detection limit of $\eta \sim 0.4$ for both reference AIBs 
considered. Resolving power helps a little, since separating single hot 
bands yields a higher peak flux in the strongest ones. 

None of these bands are detected in ISO archive data, therefore we
here obtained a direct observational upper limit on $\eta$ for neutral
naphthalene in the RR.

\begin{figure}
\begin{center}
\includegraphics[width=\hsize]{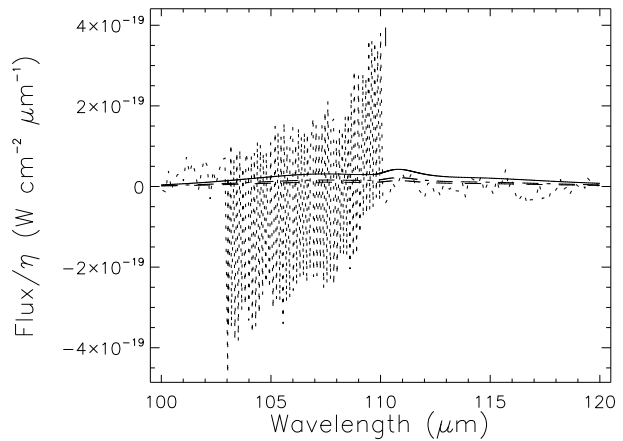}
\end{center}
\caption{Comparison between the estimated IR emission spectrum of anthracene
(C$_{14}$H$_{10}$) and an ISO spectrum of the RR in the wavelength range
100\textendash120~$\mu$m. Calculated spectra, under different assumptions
\citep[see][ for details]{mul06} are drawn in dashed, dash\textendash dotted, 
and continuous lines; the continuum\textendash subtracted ISO spectrum is shown as a 
dotted line. The central position of the fundamental of the expected 
anthracene band is marked by a tick, which shows the effect of 
anharmonic shifts.}
\label{anthracene_110}
\end{figure}

Turning to anthracene, its strongest emission bands in the RR
are again the in\textendash plane C\textendash H stretch at $\sim3.3~\mu$m and the out\textendash of\textendash plane 
C\textendash H bend at $\sim11.3$; the fraction of the total IR emission
of neutral anthracene in such bands is calculated to be respectively 37.5\%
and 17.2\%. The integrated intensities in the former of 
these bands observed in the RR was given above, the one of the $\sim11.3~\mu$m 
band is measured to be $F_\mathrm{11.3}$~=~1.39~10$^{-16}$~W~cm$^{-2}$.
Our model yields, for neutral anthracene, 
$\mathcal{P}_\mathrm{3.3}$~=~4.7~10$^{-26}$~W~sr$^{-1}$ and
$\mathcal{P}_\mathrm{11.3}$~=~2.15~10$^{-26}$~W~sr$^{-1}$, while the values
of S and $\displaystyle\frac{dF_\mathrm{cont}}{d\lambda}$ estimated from ISO 
archive data and those of 
$\displaystyle\left(\frac{d\mathcal{P}_\mathrm{ref}}{d\lambda}\right)_\mathrm{peak}$
calculated by our model (see Figs.~\ref{anthra_n_58} to \ref{anthra_n_66})
are listed in Table~\ref{pupamela2}, along with the resulting detection 
limits at $\sim$2$\sigma$ level.
\begin{table}
\caption{Detection limits estimated for three test bands of neutral 
anthracene, along with the parameters used to derive them. Two entries
are listed for the perpendicular band at 26.35~$\mu$m, corresponding to
the resolving power of SWS with and without the Fabry\textendash Perot filter.}
\label{pupamela2}
\begin{center}
\begin{tabular}{ccccccc}
\hline \hline \noalign{\smallskip} 
Band & S & $\displaystyle\frac{dF_\mathrm{cont}}{d\lambda}$ & 
R &
$\displaystyle\left(\frac{d\mathcal{P}_\mathrm{ref}}{d\lambda}\right)_\mathrm{peak}$ &
$\eta_\mathrm{3.3}$ & $\eta_\mathrm{11.3}$\\
\noalign{\smallskip} 
($\mu$m) & & $\displaystyle\left(\frac{\mathrm{W}}{\mathrm{cm^2~\mu m}}\right)$ & 
& $\displaystyle\left(\frac{\mathrm{W}}{\mathrm{sr~\mu m}}\right)$ & \\
\noalign{\smallskip} \hline \noalign{\smallskip}
21.26 & 50 & 3.1~10$^{-16}$ & \textemdash{} & 4.2~10$^{-27}$ & $>1$ & 0.44 \\
\multirow{2}*{26.35} & \multirow{2}*{90} & \multirow{2}*{1.9~10$^{-16}$} &
l & 5.1~10$^{-28}$ & $>1$ & $>1$ \\
 & & & h & 7.2~10$^{-28}$ & $>1$ & 0.86\\
43.68 & 75 & 4.9~10$^{-17}$ & \textemdash{} & 3.9~10$^{-28}$ & $>1$ & 0.52 \\
\multirow{2}*{110.23} & \multirow{2}*{25} & \multirow{2}*{1.5~10$^{-18}$} & 
l & 6.0~10$^{-29}$ & $>1$ & 0.33 \\
 & & & h & 7.2~10$^{-29}$ & $>1$ & 0.27 \\
\noalign{\smallskip} \hline 
\end{tabular}
\end{center}
\end{table}
Classical AIBs are not the only observable quantity against which 
ratios, and consequently $\eta$ values, can be computed. Neutral anthracene, 
phenanthrene and pyrene were proposed as possible carriers of the
Blue Luminescence \citep[BL, ][]{vij04,vij05}, a fluorescence 
phenomenon observed in the Red Rectangle and, subsequently, in several
other astronomical sources. In a previous work, we demonstrated that
phenanthrene and pyrene could be ruled out based on the failure to 
detect their predicted IR emission spectra \citep{mul06}. Anthracene
remained, of the three, the only candidate compatible with available
ISO observations of the Red Rectangle. The strongest constraint appeared
to be the undetected longest wavelength band at $\sim$110~$\mu$m, which 
however contained a degree of uncertainty due to the unknown band
profile. We here reproduce the comparison of the calculated band with
available observations \emph{including} the detailed modelling of band
profile. This is shown in Fig.~\ref{anthracene_110}. The line of reasoning
leading to Eq.~(\ref{etalimit}) can be identically retraced substituting 
the integrated BL flux in the place of the integrated flux in a classical 
AIB \citep[see ][ for a detailed derivation]{mul06}. This yields 
$\eta_\mathrm{BL}>1$ for anthracene, i.~e. it confirms that this band is 
undetectable with the available ISO database observations. It is however 
apparent (see Fig.~\ref{anthracene_110}) that this band is expected to 
be just slightly below the detection limit, and an increase in S/N of 
a factor of $\sim$5 ought to reveal it, if anthracene indeed produces the 
observed BL.

\section{Discussion and conclusions}\label{discussion}

The upper limits we derived are on the relative contribution of 
specific molecules to well defined classical AIBs, specifically at 
$\sim$3.3, $\sim$11.3 and $\sim12.7~\mu$m. They can be easily converted into
absolute abundance limits, but this implies an assumption on their
spatial distribution in the observed source and on the detailed 
assumed scaling of RF intensity. The terms depending on such assumptions 
cancel when using ratios of bands, making them more robust.

\begin{figure}

\includegraphics[width=\hsize]{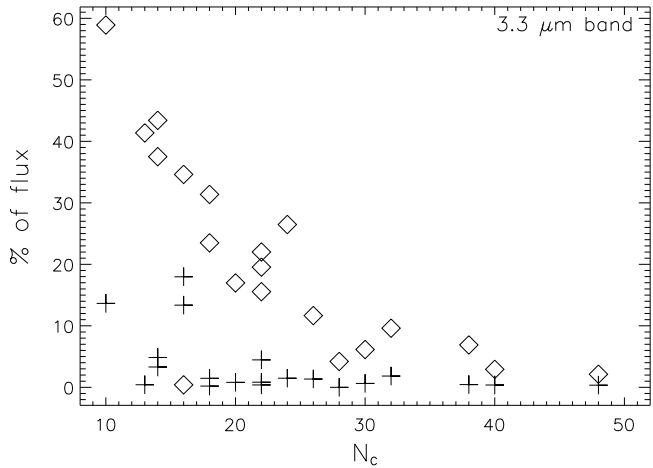}
\caption{Percentage of total IR flux emitted in the in\textendash plane C\textendash H stretch 
near $\sim3.3~\mu$m, as computed by our model for the sample of molecules in
\citet{mul06b}, as a function of molecular size.
Neutral species are represented by diamonds, cations by crosses. For each 
molecule, we here considered all bands in the range 3.2 to 3.37~$\mu$m, which
corresponds to the observed width of the $3.3~\mu$m band in the RR.}
\label{frac3_3}
\end{figure}

\begin{figure}

\includegraphics[width=\hsize]{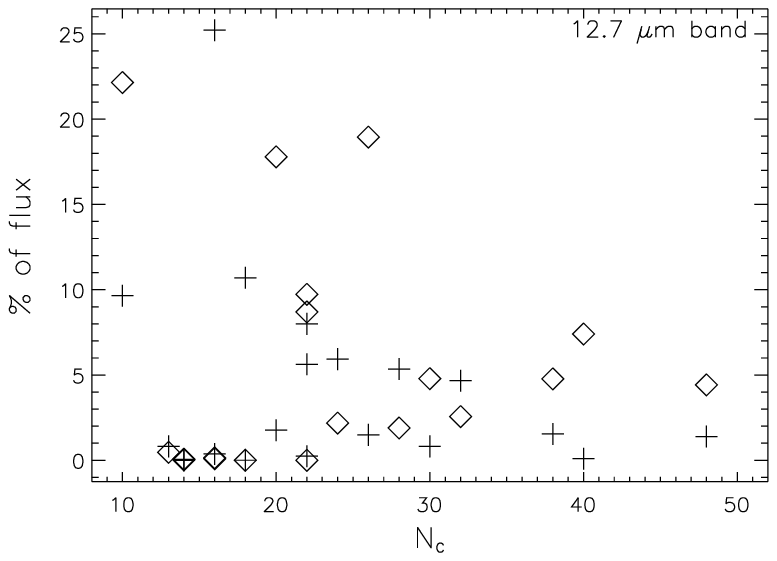}
\caption{Same as Fig.~\ref{frac3_3} for the out\textendash of\textendash plane C\textendash H bend 
at $\sim12.7~\mu$m. For each molecule, we here considered all bands in the range 
12.3 to 13.1~$\mu$m, which corresponds to the observed width of the 
$12.7~\mu$m band in the RR.}
\label{frac12_7}
\end{figure}

Despite obtaining very similar numerical values for the upper limits on
$\eta_\mathrm{3.3}$ and $\eta_\mathrm{12.7}$ for naphthalene, their relevance is 
different. The $\sim3.3~\mu$m band is produced efficiently by small neutral 
molecules; the intensity of this band is suppressed in cations and rapidly
decays with molecular size, as shown by Fig.~\ref{frac3_3}.
This implies that the PAHs significantly contributing to the $\sim3.3~\mu$m band 
are a small subset of the whole population, namely only the small ones. 
The number of different species in this subset is accordingly very 
much smaller, thus $<40\%$ of the small PAHs (i.~e. $\eta_\mathrm{3.3}$)
is a much stronger constraint than $<40\%$ of all PAHs. 
This trend is much weaker for the flux fraction in the band at $\sim12.7~\mu$m, 
as shown by Fig.~\ref{frac12_7}. This is a consequence of the well\textendash known
variation in the position of the out\textendash of\textendash plane C\textendash H bend depending on specific
molecular parameters (e.~g. \emph{solo}, \emph{duo}, \emph{trio} modes 
etc.), hence chemical diversity produces a large scatter which dominates over
size effects. This means that the band at $\sim12.7~\mu$m is not produced 
preferentially by a subset of PAHs defined by size, but instead by structure.

\begin{figure}

\includegraphics[width=\hsize]{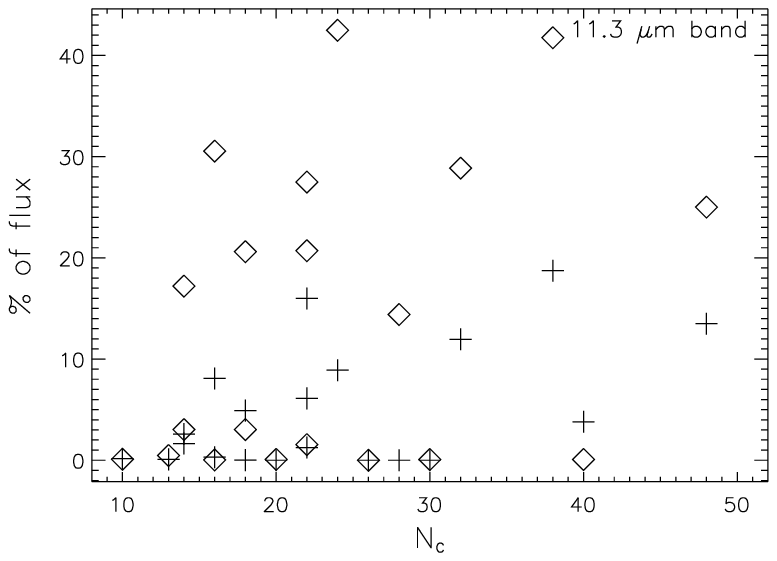}
\caption{Same as Fig.~\ref{frac3_3} for the out\textendash of\textendash plane C\textendash H bend 
at $\sim11.3~\mu$m. For each molecule, we here considered all bands in the range 
10.9 to 11.7~$\mu$m, which corresponds to the observed width of the 
$11.3~\mu$m band in the RR.}
\label{frac11_3}
\end{figure}

As to anthracene, we obtained rather weak constraints for $\eta_\mathrm{3.3}$, 
somewhat stronger constraints on $\eta_\mathrm{11.3}$. However, 
Fig.~\ref{frac11_3} shows that larger molecules can be expected to contribute
a considerable fraction of the flux emitted in this band, which means that 
$\sim$40\% is not such a stringent limit for $\eta_\mathrm{11.3}$. As for
the band at $\sim12.7~\mu$m, the band at $\sim11.3~\mu$m is produced preferentially
by a subset of molecules defined primarily by structure, more than by size.

While we applied our procedure to neutral naphthalene and anthracene, it can 
be generalised to any PAH for which our model is applicable and to any
astronomical source for which observations of classical AIBs (or some other
suitable tracer of PAHs) are available. As already mentioned, \emph{all} 
PAHs have butterfly IR\textendash active modes, which usually are the lowest frequency 
modes of each molecule and give rise to perpendicular bands \citep{mul06b}. 
The lowest frequency vibrational modes are almost always well separated 
in energy (except for highly symmetric molecules), meaning that strong 
Coriolis perturbations are not likely to occur for them and they ought 
to display Q branches which can be very sharp, since the effective momenta 
of inertia can be very close between the upper and lower
states of the transition. In turn, according to the present calculations
these Q branches ought to contain $\sim$20\% of the total flux in the band, 
the distribution among fundamental and hot bands depending on the detailed 
statistics of the process and, essentially, on the vibrational modes which 
are left in an excited state upon decoupling. 
The spacing between the fundamental and various hot bands depends critically 
on the anharmonic vibrational constants $\chi_{ij}$, which vary with the specific 
molecule and band considered. 
If a perpendicular band is emitted preferentially when the molecule 
has a very low excitation energy, then relatively few hot bands will 
contribute to it, their Q branches remaining well resolved and
sharp, and the P and R branches blending and being somewhat blurred
as a result. The same will happen if a few $\chi_{ij}$ are much larger than the 
others, producing a well\textendash defined sequence of sharp Q branches.

We here modelled in detail the profiles of one perpendicular band of 
naphthalene and three of anthracene. Among them, three display sharp
Q branches (Figs.~\ref{naphtha_n_48}, \ref{anthra_n_61} and 
\ref{anthra_n_66}), in which fundamental and hot bands can be separated 
with adequate resolving power, the other one (Fig.~\ref{anthra_n_58}) 
is red shaded to the point of partially washing out its structure. 
In this very limited sample, therefore, most perpendicular bands 
possess a combination of molecular parameters which yields the maximum 
spectral contrast. On a large sample of molecules, it can be reasonably 
expected that this will happen for some intense bands, producing the most 
favorable conditions for their detection.

In general, the average 
rotational energy of the molecule is expected to scale roughly with the 
average energy of the vibrational photons emitted \cite[see e.~g.][]{rou97}, 
which implies that larger molecules ought to have a lower rotational energy 
which, together with decreasing rotational constants, in turn translates 
to narrower rotational profiles and better detectability.

The mixture of PAHs in space 
probably contains a wide variety of different molecules, in principle in 
many different hydrogenation and ionisation states. However, 
\citet{lep01,lep03} showed that in any given environment 
the population of a given PAH is expected to be dominated by 
one or two ionisation and hydrogenation states, essentially depending on
its size and physical conditions. Furthermore, chemical selection effects 
will favour the most stable species over the others, so that at least 
\emph{some} species may be abundant enough to exceed their detection
limit on \emph{some} $\eta_\mathrm{AIB}$ (or some other PAH tracer).
Moreover, such a detection limit will be pushed one or
two orders of magnitude lower by the much higher sensitivity of the 
instruments on board the forthcoming Herschel Space Observatory and by the
much reduced dust continuum emission at longer wavelengths.

A systematic study of PAH band profiles and IR flux ratios is necessary 
to select the most promising diagnostics and species for identification.
The search for single, specific PAHs in the far\textendash IR is a challenging, 
but promising task.

\begin{acknowledgements}
G.~Malloci acknowledges the ``Minist\`ere de la Recherche'' and G.~Mulas
the CNRS for the financial support during their stay at CESR in Toulouse.
We thank Aude Simon for helping us to effectively use the 
\textsc{Gaussian03} package. Part of the calculations used here were 
performed using CINECA and CALMIP supercomputing facilities.
This work was supported by  the European Research Training Network
``Molecular Universe'' (MRTN-CT-2004-512302).
\end{acknowledgements}


\appendix

\section{Computed molecular parameters from the vibro\textendash rotational 
analyses}

In the following we report some of the molecular parameters obtained in the
vibro\textendash rotational analyses performed for neutral naphthalene and neutral
anthracene at the B3LYP/\mbox{4\textendash31G} level of theory. 
Tables~\ref{chinaphthalene} and \ref{chianthracene}
report, respectively, the vibrational anharmonic constants $\chi_{ij}$ for the 
three bands of neutral naphthalene and the four bands opf neutral anthracene 
considered. The changes of the effective rotational constants as a function of
vibrational state are reported in Tables~\ref{vibrot_alpha_naphthalene} for
naphthalene and \ref{vibrot_alpha_anthracene} for anthracene. For consistency,
we list in the tables the frequencies we previously obtained \citep{mul06b}
using the \textsc{NWChem} package \citep{str03} at the same level of theory.
Frequencies obtained by \textsc{Gaussian03}  \citep{g03} are coincident within 
numerical errors. We give more significant digits than the actually expected 
accuracy to distinguish very close vibrational modes.

\begin{table}[h!]
\begin{center}
\caption{Anharmonic constants $\chi_{ij}$ (expressed in cm$^{-1}$) obtained at the 
B3LYP/4\textendash31G level for the three vibrational modes of naphthalene 
neutral for which we modelled the detailed rotational structure. }
\label{chinaphthalene}
\begin{tabular}{cccc}
\hline \hline
\noalign{\smallskip}
Fundamental & 15.820~$\mu$m & 27.731~$\mu$m & 58.196~$\mu$m \\
($\mu$m) & \multicolumn{3}{c}{(cm$^{-1}$)} \\
\noalign{\smallskip}
\hline
\noalign{\smallskip}
3.249 & -0.126 & 0.004 & 0.035 \\
3.250 & -0.123 & -0.018 & 0.039 \\
3.265 & -0.198 & 0.083 & 0.012 \\
3.267 & -0.198 & 0.032 & 0.018 \\
3.280 & -0.156 & 0.217 & -0.014 \\
3.284 & -0.130 & 0.175 & -0.026 \\
3.286 & -0.198 & 0.131 & 0.002 \\
3.288 & -0.178 & 0.123 & -0.008 \\
6.179 & -1.360 & -1.051 & -0.449 \\
6.278 & -0.406 & -1.057 & -0.306 \\
6.407 & -1.347 & -0.198 & -0.239 \\
6.632 & -1.163 & -0.519 & -0.309 \\
6.836 & -0.735 & 0.180 & -0.077 \\
6.842 & -0.659 & -0.563 & -0.240 \\
7.146 & -0.486 & 0.061 & -0.055 \\
7.335 & -1.522 & -0.039 & -0.332 \\
7.364 & -1.765 & 0.820 & -0.110 \\
7.875 & -0.630 & 0.704 & -0.110 \\
7.961 & -0.305 & 0.038 & -0.065 \\
8.273 & -0.406 & -1.356 & -0.327 \\
8.527 & 0.011 & 1.270 & 0.207 \\
8.551 & -0.344 & -0.482 & -0.010 \\
8.658 & 0.044 & 0.165 & -0.055 \\
8.838 & -0.382 & 0.441 & -0.229 \\
9.812 & 0.082 & 0.824 & -0.034 \\
9.906 & 0.089 & 0.099 & -0.066 \\
10.065 & -0.189 & -0.052 & -0.021 \\
10.141 & -0.140 & -0.115 & -0.126 \\
10.389 & -0.089 & -0.004 & -0.294 \\
10.601 & -0.024 & -0.068 & -0.157 \\
10.638 & -1.126 & 0.150 & -0.083 \\
11.294 & -0.221 & -0.088 & -0.457 \\
11.937 & -0.325 & -0.089 & -0.245 \\
12.545 & -0.088 & -1.264 & 0.292 \\
12.691 & -0.064 & -0.176 & -0.155 \\
12.957 & -0.335 & 0.050 & -0.190 \\
13.312 & 0.859 & -0.252 & -0.077 \\
13.937 & 0.046 & 0.087 & -0.093 \\
15.825 & 0.105 & -0.104 & -0.200 \\
15.895 & -0.226 & -0.221 & -0.463 \\
19.503 & 0.057 & -0.095 & 0.339 \\
19.536 & 0.050 & 0.254 & 0.807 \\
20.878 & 0.237 & 0.226 & -0.054 \\
21.162 & -0.225 & -0.324 & -0.349 \\
25.745 & 0.259 & -0.295 & -0.579 \\
27.737 & -0.104 & 0.236 & 0.363 \\
53.950 & -0.113 & -0.460 & -0.230 \\
58.565 & -0.200 & 0.363 & -0.258 \\
\noalign{\smallskip}
\hline
\end{tabular}
\end{center}
\end{table}

\begin{table}[h!]
\begin{center}
\caption{The $\mathrm{a}_i$,  $\mathrm{b}_i$ and $\mathrm{c}_i$ constants 
(in units of 10$^{-4}$ cm$^{-1}$) as obtained at the B3LYP/4\textendash31G level for 
neutral naphthalene.
These quantities express the change of the rotational constants as a function 
of the vibrational mode. For comparison, the rotational constants A, B and C 
in the vibrational ground state are 0.103289, 0.040779 and 0.029241 cm$^{-1}$, 
respectively.}
\label{vibrot_alpha_naphthalene}
\begin{tabular}{cccc}
\hline \hline
\noalign{\smallskip}
Fundamental & a$_i$ & b$_i$ & c$_i$ \\
($\mu$m) & \multicolumn{3}{c}{(10$^{-4}$ cm$^{-1}$)} \\
\noalign{\smallskip}
\hline
\noalign{\smallskip}
3.249 & 0.346 & 0.080 & 0.065 \\
3.250 & 0.347 & 0.077 & 0.063 \\
3.265 & 0.320 & 0.087 & 0.068 \\
3.267 & 0.323 & 0.081 & 0.064 \\
3.280 & 0.325 & 0.096 & 0.073 \\
3.284 & 0.323 & 0.093 & 0.070 \\
3.286 & 0.313 & 0.087 & 0.069 \\
3.288 & 0.313 & 0.088 & 0.068 \\
6.179 & 0.800 & 0.519 & 0.136 \\
6.278 & 0.471 & 0.387 & 0.187 \\
6.407 & 1.061 & 0.307 & 0.478 \\
6.632 & 0.661 & 0.241 & 0.303 \\
6.836 & 0.410 & 0.055 & -2.109 \\
6.842 & -0.124 & 0.302 & 2.421 \\
7.146 & -0.199 & 0.136 & 0.081 \\
7.335 & 1.918 & 0.531 & 0.395 \\
7.364 & 0.683 & 0.339 & 0.280 \\
7.875 & 0.299 & 0.161 & 0.104 \\
7.961 & 0.214 & 0.094 & 0.220 \\
8.273 & -0.055 & 0.458 & 0.324 \\
8.527 & -0.625 & 0.003 & 0.077 \\
8.551 & 0.250 & 0.098 & 0.028 \\
8.658 & -0.199 & 0.049 & 0.069 \\
8.838 & -0.488 & 0.102 & 0.324 \\
9.812 & 0.809 & -0.042 & 0.144 \\
9.906 & 0.782 & 0.009 & 0.112 \\
10.065 & 0.916 & 0.163 & 0.022 \\
10.141 & 0.954 & 0.153 & 0.017 \\
10.389 & 0.448 & 0.192 & 0.012 \\
10.601 & -2.638 & 0.159 & 0.002 \\
10.638 & 2.807 & 0.004 & 0.141 \\
11.294 & 0.546 & 0.157 & 0.005 \\
11.937 & 0.194 & 0.017 & -0.001 \\
12.545 & -2.053 & 0.193 & 0.117 \\
12.691 & 2.656 & 0.057 & -0.013 \\
12.957 & 0.333 & 0.085 & -0.029 \\
13.312 & 0.514 & 0.057 & 0.139 \\
13.937 & 0.319 & 0.107 & -0.020 \\
15.825 & -40.137 & -0.023 & 0.113 \\
15.895 & 40.437 & 0.062 & -0.045 \\
19.503 & -0.220 & -0.037 & -11.605 \\
19.536 & -0.220 & 0.061 & 11.782 \\
20.878 & 0.053 & 0.025 & -0.097 \\
21.162 & 0.692 & -0.008 & -0.104 \\
25.745 & 0.733 & -0.042 & -0.113 \\
27.737 & -1.792 & -0.050 & 0.069 \\
53.950 & 0.339 & 0.146 & -0.150 \\
58.565 & 2.085 & -0.262 & -0.249 \\
\noalign{\smallskip}
\hline
\end{tabular}
\end{center}
\end{table}

\clearpage

\begin{table*}
\begin{center}
\caption{Same as Fig.~\ref{chinaphthalene} for the four vibrational 
modes of neutral anthracene for which we modelled the detailed rotational 
structure.}\label{chianthracene}
\begin{tabular}{ccccc|ccccc}
\hline \hline
\noalign{\smallskip}
Fund. & 21.232~$\mu$m & 26.336~$\mu$m & 43.060~$\mu$m & 109.937~$\mu$m &
Fund. & 21.232~$\mu$m & 26.336~$\mu$m & 43.060~$\mu$m & 109.937~$\mu$m \\
($\mu$m) & \multicolumn{4}{c}{(cm$^{-1}$)} &
($\mu$m) & \multicolumn{4}{c}{(cm$^{-1}$)} \\
\noalign{\smallskip}
\hline\noalign{\smallskip}
\noalign{\smallskip}
3.249 & -0.079 & -0.063 & -0.027 & 0.045 & 9.946 & 0.062 & -0.100 & 0.121 & -0.041 \\
3.249 & -0.078 & -0.065 & -0.030 & 0.044 & 10.102 & -0.389 & -0.628 & -0.032 & -0.037 \\
3.263 & -0.096 & -0.033 & -0.004 & 0.029 & 10.116 & -0.275 & -0.742 & -0.035 & -0.036 \\
3.263 & -0.095 & -0.035 & -0.007 & 0.030 & 10.407 & -0.660 & -0.563 & -0.009 & -0.055 \\
3.282 & -0.149 & 0.044 & 0.058 & 0.007 & 10.470 & -0.372 & -0.520 & -0.008 & -0.059 \\
3.283 & -0.128 & 0.039 & 0.042 & 0.005 & 10.875 & -0.420 & 0.020 & -0.284 & 0.007 \\
3.284 & -0.135 & 0.023 & 0.035 & 0.014 & 11.003 & -1.283 & -1.905 & -0.057 & -0.163 \\
3.286 & -0.110 & 0.011 & 0.018 & 0.017 & 11.005 & -0.348 & -0.140 & -0.148 & 0.205 \\
3.292 & -0.317 & 0.085 & 0.086 & -0.008 & 11.315 & -2.030 & -0.563 & -0.161 & -0.036 \\
3.295 & -0.299 & 0.087 & 0.081 & -0.011 & 11.690 & -0.356 & -0.376 & -0.030 & -0.100 \\
6.162 & -0.331 & -0.554 & -0.338 & -0.143 & 11.982 & -1.000 & -0.666 & -0.035 & -0.096 \\
6.183 & -0.703 & -0.614 & 1.810 & -0.184 & 12.563 & -0.517 & -0.324 & 0.328 & 0.008 \\
6.329 & -0.421 & -0.520 & -1.023 & -0.177 & 13.043 & -3.945 & -0.560 & 0.021 & 0.197 \\
6.486 & -0.559 & -0.572 & -0.028 & -0.119 & 13.173 & -0.291 & -3.714 & 0.032 & 0.039 \\
6.514 & -0.376 & -0.471 & 0.081 & -0.161 & 13.434 & -0.667 & 0.317 & 0.023 & -0.098 \\
6.736 & -0.137 & -0.297 & -0.128 & -0.096 & 13.512 & -0.712 & -0.806 & -0.017 & -0.027 \\
6.860 & -0.295 & 0.048 & -0.554 & -0.126 & 13.709 & -0.367 & -0.668 & 0.031 & 0.035 \\
6.864 & -0.021 & -0.365 & 0.243 & -0.040 & 15.329 & -0.554 & -0.136 & -0.012 & -0.209 \\
7.147 & -0.144 & 0.118 & -0.072 & -0.031 & 15.706 & 0.187 & 0.120 & 0.016 & -0.160 \\
7.218 & -0.192 & -0.813 & -0.453 & -0.110 & 16.336 & 0.066 & -0.027 & -0.481 & 0.131 \\
7.218 & -0.829 & -0.933 & 0.704 & -0.098 & 17.109 & -0.112 & -0.470 & -0.277 & -0.088 \\
7.434 & -0.989 & -0.699 & -0.845 & -0.140 & 18.710 & -0.287 & -0.037 & -0.073 & 0.265 \\
7.605 & -0.008 & 1.095 & 0.019 & -0.067 & 20.037 & -0.060 & -0.499 & 0.050 & 0.151 \\
7.768 & 0.290 & -0.460 & 0.512 & -0.031 & 21.019 & 0.209 & -0.171 & 0.016 & -0.192 \\
7.848 & 2.118 & -0.500 & -0.024 & -0.087 & 21.263 & 0.040 & -0.065 & -0.030 & -0.002 \\
7.899 & -0.355 & -0.226 & 0.006 & -0.135 & 25.448 & -0.200 & -0.338 & 0.664 & -0.424 \\
8.318 & -0.272 & -0.491 & 0.045 & -0.002 & 25.625 & 0.109 & -0.064 & -0.244 & 0.915 \\
8.473 & -0.318 & -0.152 & 0.056 & -0.137 & 26.349 & -0.065 & 0.171 & -0.271 & -0.181 \\
8.553 & -0.196 & 0.349 & 0.069 & -0.060 & 37.412 & -0.237 & -0.220 & -0.408 & 0.017 \\
8.623 & -0.535 & 2.072 & -0.178 & -0.076 & 43.009 & -0.129 & -0.554 & -0.978 & -0.281 \\
8.668 & -0.349 & -0.198 & -0.221 & -0.037 & 43.681 & -0.030 & -0.271 & 0.024 & -0.587 \\
9.107 & -0.588 & -1.000 & -1.709 & -0.117 & 82.090 & -0.261 & -0.269 & -0.828 & -0.173 \\
9.911 & 0.105 & -0.118 & 0.200 & -0.040 & 110.231 & -0.002 & -0.181 & -0.587 & -0.146 \\
\hline
\end{tabular}
\end{center}
\end{table*}

\clearpage

\begin{table*}
\begin{center}
\caption{Same as Fig.~\ref{vibrot_alpha_naphthalene} for neutral anthracene.
The rotational constants A, B and C in the vibrational ground state are 
0.071063, 0.014959 and 0.012361 cm$^{-1}$, respectively.}
\label{vibrot_alpha_anthracene}
\begin{tabular}{cccc|cccc}
\hline \hline
\noalign{\smallskip}
Fund. & a$_i$ & b$_i$ & c$_i$ & Fund. & a$_i$ & b$_i$ & c$_i$ \\
($\mu$m) & \multicolumn{3}{c}{(10$^{-4}$ cm$^{-1}$)} & ($\mu$m) & \multicolumn{3}{c}{(10$^{-4}$ cm$^{-1}$)} \\
\noalign{\smallskip}
\hline
\noalign{\smallskip}
3.249 & 0.166 & 0.014 & 0.014 & 9.946 & 0.433 & -0.001 & 0.027 \\
3.249 & 0.166 & 0.014 & 0.014 & 10.102 & 0.430 & 0.025 & 0.005 \\
3.263 & 0.155 & 0.015 & 0.015 & 10.116 & 0.434 & 0.024 & 0.005 \\
3.263 & 0.155 & 0.015 & 0.014 & 10.407 & 0.193 & 0.029 & 0.003 \\
3.282 & 0.161 & 0.020 & 0.018 & 10.470 & 0.072 & 0.023 & 0.001 \\
3.283 & 0.157 & 0.018 & 0.016 & 10.875 & 0.154 & 0.018 & 0.035 \\
3.284 & 0.155 & 0.018 & 0.017 & 11.003 & 0.211 & 0.031 & 0.004 \\
3.286 & 0.152 & 0.016 & 0.015 & 11.005 & -0.323 & 0.009 & 0.024 \\
3.292 & 0.176 & 0.029 & 0.024 & 11.315 & 0.401 & 0.016 & 0.000 \\
3.295 & 0.172 & 0.029 & 0.024 & 11.690 & 0.199 & 0.026 & 0.002 \\
6.162 & 0.296 & 0.106 & 0.069 & 11.982 & 0.101 & 0.024 & 0.001 \\
6.183 & 0.375 & 0.147 & 0.084 & 12.563 & 0.261 & 0.052 & 0.059 \\
6.329 & 0.206 & 0.122 & 0.082 & 13.043 & 0.141 & 0.013 & -0.005 \\
6.486 & 0.539 & 0.081 & 0.106 & 13.173 & 0.238 & 0.008 & -0.002 \\
6.514 & 0.356 & 0.086 & 0.087 & 13.434 & 0.223 & -0.004 & -0.010 \\
6.736 & 0.276 & 0.055 & 0.063 & 13.512 & 0.313 & 0.016 & 0.024 \\
6.860 & -0.009 & 0.092 & -0.239 & 13.709 & 0.092 & 0.014 & -0.002 \\
6.864 & 0.198 & 0.024 & 0.334 & 15.329 & 0.008 & 0.037 & -0.033 \\
7.147 & -0.073 & 0.036 & 0.027 & 15.706 & -1.289 & -0.003 & 0.015 \\
7.218 & 0.671 & 0.117 & 0.117 & 16.336 & -0.177 & 0.011 & 0.090 \\
7.218 & 0.784 & 0.089 & 0.079 & 17.109 & 1.376 & -0.018 & -0.015 \\
7.434 & 0.409 & 0.136 & 0.096 & 18.710 & -0.250 & 0.015 & 0.022 \\
7.605 & -0.010 & 0.095 & 0.087 & 20.037 & 0.424 & -0.019 & -0.028 \\
7.768 & 0.111 & 0.028 & 0.000 & 21.019 & 0.194 & -0.007 & -0.023 \\
7.848 & 0.082 & 0.052 & 0.062 & 21.263 & 0.013 & -0.004 & -0.023 \\
7.899 & 0.283 & 0.136 & 0.142 & 25.448 & -0.683 & -0.002 & -1.019 \\
8.318 & 0.165 & 0.009 & 0.031 & 25.625 & -0.012 & 0.020 & 1.062 \\
8.473 & -0.268 & 0.004 & 0.014 & 26.349 & 0.355 & -0.023 & -0.032 \\
8.553 & -0.180 & 0.035 & 0.037 & 37.412 & 0.261 & -0.006 & -0.034 \\
8.623 & 0.328 & 0.071 & -0.448 & 43.009 & 0.904 & -0.060 & -0.059 \\
8.668 & -0.162 & 0.017 & 0.565 & 43.681 & -1.292 & -0.026 & 0.009 \\
9.107 & -0.236 & 0.070 & 0.055 & 82.090 & 0.124 & 0.026 & -0.042 \\
9.911 & 0.391 & 0.002 & 0.030 & 110.231 & 1.647 & -0.082 & -0.091 \\
\noalign{\smallskip}
\hline
\end{tabular}
\end{center}
\end{table*}

\end{document}